# Signature of growth-deposition technique on the properties of PECVD and thermal SiO$_2$

**Subimal Majee[1]\*, Devesh Barshilia[1], Sanjeev Kumar[1], Prabhash Mishra[2] and Jamil Akhtar[1]**
[1]CSIR-Central Electronics Engineering Research Institute (CEERI), Pilani 333031, Rajasthan, India
[2]Nano-Science Center, Jamia Millia Islamia, New Delhi, 110025, India
\*E-mail: subimal.majee@polytechnique.edu

**Abstract.** In this article, we report the process induced variation in the characteristics of PECVD deposited and thermally grown silicon dioxide (SiO$_2$) thin film. We find key differences in the porosity, arrangement of the nano-pores, surface roughness, refractive index and electrical resistivity of the SiO$_2$ thin films obtained by the two methods. While the occurrence of the nanoporous structure is an inherent property of the material and independent of the process of film growth or deposition, the arrangements of these nano-pores in the oxide film is process dependent. The distinct arrangements of the nano-pores are signatures of the deposition/growth processes. Morphological analysis has been carried out to demonstrate the difference between oxides either grown by thermal oxidation or through PECVD deposition. The tunable conductive behavior of the metal filled nano-porous oxides is also demonstrated, which has potential to be used as conductive oxides in various applications.

Keywords: Silicon dioxide; PECVD; Thermal oxidation; Porosity

## 1. Introduction

Nanoporosity in thin membranes and thin films is either an inherent property of the materials or sometimes process dependent. Although, porosity is mostly undesirable in active devices, there are several areas where nano-porous membranes or thin films are absolutely necessary. As a matter of fact, each of the nano-pores inside a porous material has the potential to be an active device if properly tuned by some means. It is still an open field of research the communications between each of the nano-pores in nano-scale dimensions and their effect as a whole in the bulk material. In recent times, nano-porous thin films have attracted much research interests owing to their potential applications in fuel cells [1-6]; pressure and gas sensors [7-8]. The utilization of nanoporous membranes in such devices require geometrically controllable nanopores with controlled pore size and pore-density arrangements [6].

In the flexible electronics research, the most commonly used insulators are low temperature plasma enhanced chemical vapor deposited (PECVD) silicon dioxide (SiO$_2$) and silicon nitrides (SiN$_x$) [9]. There is a huge ongoing research on fabrication of low cost transparent conductive oxide (TCO) films [10], which has applications in flexible displays, organic solar cells, and organic light emitting diodes and so on [10-13]. Indium tin oxide (ITO) being one of the major TCOs has drawback, like, high processing cost. Therefore, alternative solutions are being researched to find out novel conductive oxide films. Silicon dioxide ($SiO_2$) thin film is one of the widely researched materials for several decades which have applications as the dielectric interlayer in the microelectronic industries [14]; moisture permeation barriers [15]; optical coatings [16] and so on. Although this material has a huge prospect in the above mentioned areas, the insulating nature of the oxide films does not allow their applications as TCOs. Expensive chemical doping is required to obtain electrical conductivity in the $SiO_2$ thin films.

The aim of this present study is to characterize the nano-porosity of thermally grown and PECVD deposited $SiO_2$ thin films. The distinct arrangements of the nano-porous matrices in those two processes show the signatures of growth/deposition methods. We also demonstrate a simple low cost method to fabricate conductive nano-porous $SiO_2$ thin films without any intentional chemical doping. The conductive nanoporous oxide films have electrical resistivity 6 orders of magnitude lower compared to the insulating $SiO_2$ films.

## 2. Experimental details
*2.1. Oxide growth in oxidation furnace and RF PECVD deposition*
Thermally grown $SiO_2$ thin films with thickness around 1 μm have been grown in oxidation furnace at CEERI-Pilani with temperature ~1100 °C. Dry-wet-dry oxidation cycles have been used for total duration of ~3 hr on cleaned Si wafer substrates. For the RF PECVD deposition of $SiO_2$ thin films (thickness ~1μm) on cleaned Si wafer substrates, the working pressure and substrate temperature have been fixed at 0.3 Torr and 300 °C. We used an optimized recipe for oxide deposition where the flow rates for the source gases were: 5 sccm ($SiH_4$); 350 sccm ($N_2O$); 300 sccm ($N_2$) and 170 sccm (Ar). The RF (13.56 MHz) power density has been fixed at 155 mW/cm2. The optimized deposition rate ($r_d$) was 50 nm/min.

*2.2. Filling of nano-pores with metal*
In order to obtain conducting oxide films, the oxide nano-porous structures are filled with Titanium (200 Å) and gold (2000 Å) in a high vacuum ($10^{-7}$ Torr) e-beam evaporation system. Gold is deposited on top of Ti in order to avoid immediate oxidation of Ti. Gold is removed from the surface using standard gold etchant and Ti is etched carefully from the surface using diluted HF solution at room temperature where the concentration of the solution remains constant.

*2.3. Characterizations of the samples*
Thickness of the $SiO_2$ thin films has been assessed through single wavelength elipsometric measurements and re-confirmed by Dektak 150 profilometer measurements. Morphological characterizations have been carried out using tapping mode atomic force microscopy (AFM) and Field Emission Scanning Electron Microscopy (FE-SEM). Electrical measurements have been carried out using a Keithley probe station.

## 3. Results and discussions

Figure 1(a-b) shows the AFM morphological images for the RF PECVD deposited and thermally grown $SiO_2$ thin films. Distinct nano-porous patterns are observed in both cases. We observe specific arrangements of the patterns of the nano-porous structures which are distinctly different from each other. While the PECVD deposited films contain clusters of grains, the thermally grown oxide exhibits a different feature. These distinctly different matrices of those oxide films deposited/grown by two completely different methods are signatures for those processes. To the best of our knowledge, we report for the first time, the unique signatures of the process dependent nature of the oxide films in terms of its nano-porosity.

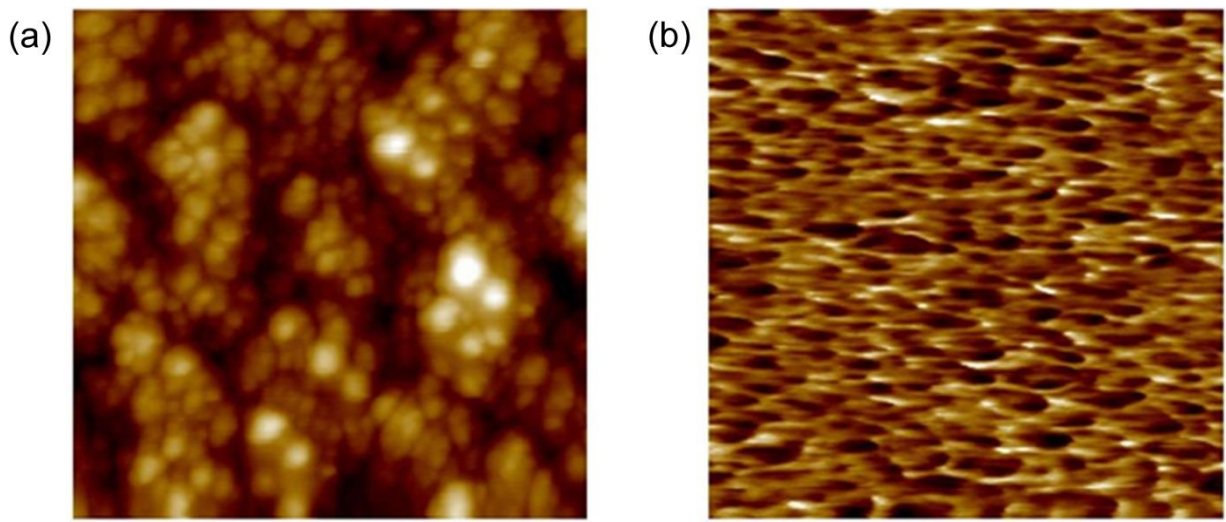

**FIGURE 1.** AFM morphological images (500 nm × 500 nm scan size) of the nano-porosity in both (a) PECVD and (b) thermally grown oxide films.

Figure 2 shows the representative FE-SEM image of the thermally grown oxide film. Nano-porosity is also observed through FE-SEM measurements. The nano-porous structures have been analyzed through Nanoscope analysis software. The pore sizes are measured and the histograms of the pore sizes are provided in Figure 3(a-b). The PECVD deposited $SiO_2$ thin films exhibit a larger variations in pore size with peak at 30 nm, whereas, a Gaussian size distribution is observed for the nano-pores in the thermally grown oxide thin films. It is interesting to note that peak positions for both cases are around 30 nm. From these observations, it is to be concluded that although the formations of nano-porous matrices are process dependent, the occurrence of nano-porosity is an inherent property for the oxide thin films.

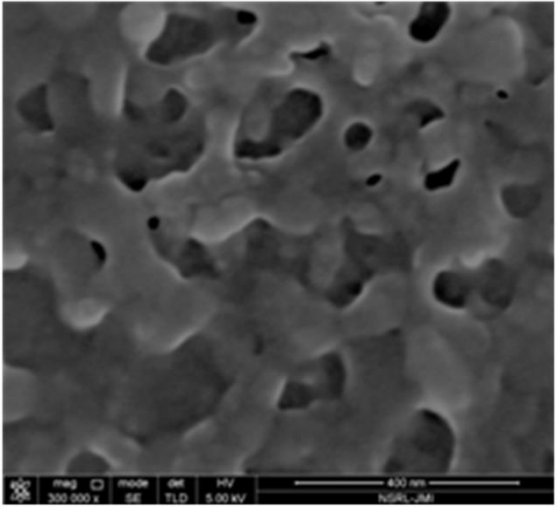

**FIGURE 2.** FE-SEM image of the nano-porous thermally grown oxide film. (scale bar 400 nm).

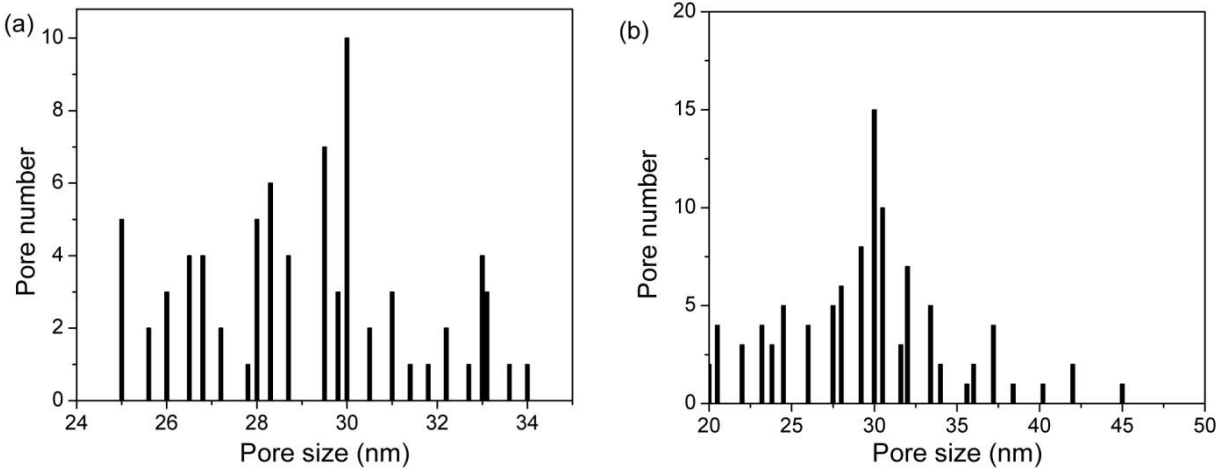

**FIGURE 3.** Histograms for the pore sizes of (a) PECVD deposited and (b) Thermally grown oxides.

Figure 4 shows the nano-porosity (*P%*) estimations for the thermally grown and PECVD deposited oxide thin films. The porosity of the oxide layer is defined by the fraction of voids within the oxide layer and can be easily determined through AFM morphological measurements, as described by F. Alfeel et al. [17]. Assuming the nano-pores to be cylindrical like nano-structures, we can estimate the nano-porosity of the films using the pore depth, number of pores per unit area and pore sizes. The estimated values of the nano-porosities are 3% for thermally grown oxides and 25% for PECVD deposited oxide films.

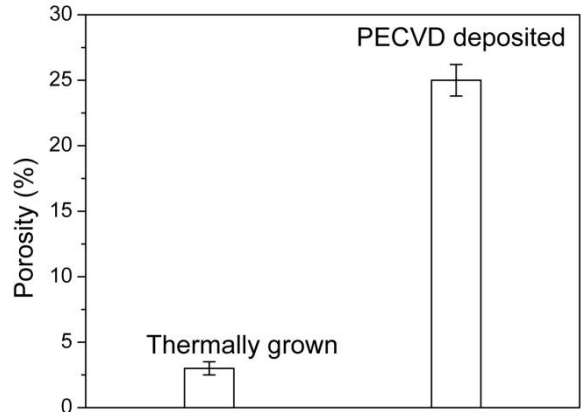

**FIGURE 4.** Estimated porosity for the thermally grown and PECVD deposited oxide thin films.

The refractive indexes of the nano-porous structures have been assessed from the nano-porosity estimations [18]. With increased nano-porosity, the density of the oxide films becomes lower. When the pore distribution is homogeneous and the pore sizes are substantially smaller than the wavelength, the nano-porosity is related to the refractive index of the porous materials ($n_P$) by the following relation:

$$P(\%) = \left[1 - \frac{(n_P^2 - 1)}{(n_d^2 - 1)}\right] \times 100$$

Where, $n_d$ is the refractive index of the non-porous material. The estimated refractive indexes for the thermally grown and PECVD deposited oxides are 1.44 and 1.36, respectively, as shown in Figure 5. The observations can be explained in terms of hydrogen (H) contents in each of the oxide films. While thermally grown oxides contains negligible amount of atomic H, the H-content in the low temperature PECVD deposited thin films is quite high (> 20 At.%) [15]. Increased amount of H-content inside the PECVD deposited oxides leads to the lower density films and lower refractive index value compared to the denser thermally grown oxides.

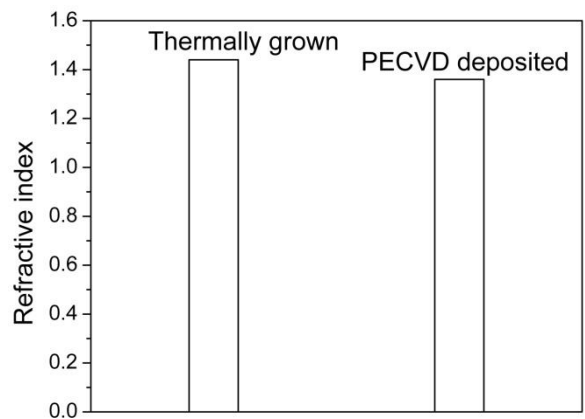

**FIGURE 5.** Calculated refractive indexes for the thermally grown and PECVD deposited oxide thin films.

The average surface roughness ($R_a$) values for both thermally grown and PECVD deposited oxides are shown in Figure 6. While we observe a smooth surface for the thermally grown oxide film, the surface roughness for the PECVD deposited films is higher. The higher surface roughness in PECVD deposited samples is likely due to the pump-down of the CVD chamber (turbulent flows pick up the dust from the surface walls of the chamber and redistribute it across the surfaces in the chamber).

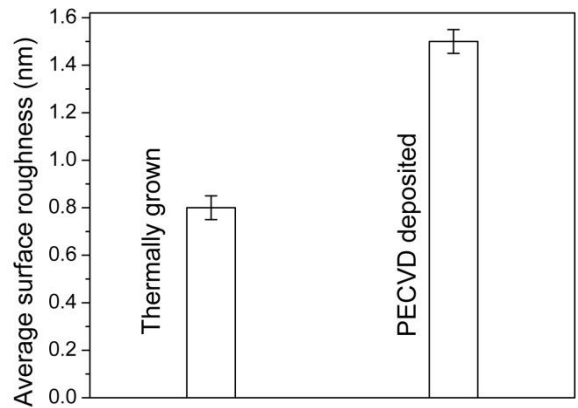

**FIGURE 6.** Average surface roughness for the thermally grown and PECVD deposited oxide thin films.

When the oxide nano-porous structures are filled with metals, we are able to tune the electrical conductivity of the oxide films. After complete removal of the metals from the oxide surfaces, only the nano-porous structures are filled with metals. Electrical characterizations are carried out before and after filling up the nano-pores with metals. Figure 7 shows the electrical resistivity measurements for metal filled nano-porous oxide and as-grown/as-deposited nano-porous oxide films. We observe 6 orders of magnitude reduction in the electrical resistivity after filling up the nano-pores with metals. The conductive $SiO_2$ thin film has the potential to be utilized as an alternative low cost TCO and as membranes for the pressure and gas sensors.

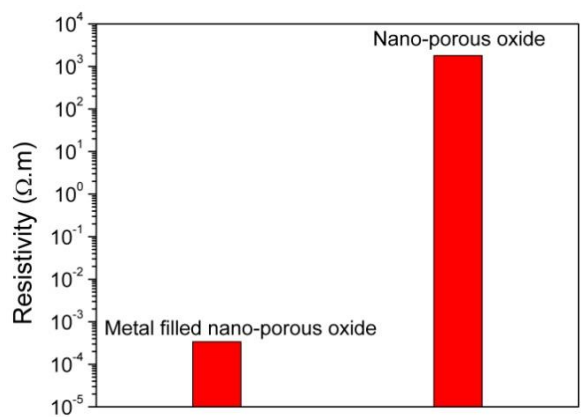

**FIGURE 7.** Resistivity comparison between metal filled nano-porous oxide and nano-porous oxide.

4. **Conclusions**

In summary, we have demonstrated process dependent arrangements of the nano-porous structures inside silicon dioxide thin films. We observe two completely different types of nano-porous patterns for thermally grown and PECVD deposited oxide thin films, which are signatures for those two processes. Morphological measurements confirm the average pore sizes in those two cases are about 30 nm which confirms that formation of pores inside $SiO_2$ is independent of the deposition/growth processes. Without any intentional doping, we are also able to tune the electrical conductivity of the nano-porous oxide films by simply filling up the nano-pores. This simple and low cost process enables fabrication of conductive oxide films. The conductive $SiO_2$ films can be used as the base materials for the next generation flexible smart systems.

## 5. Acknowledgments

Prof. Santanu Chaudhury, Director CEERI-Pilani is thankfully acknowledged for his support. The authors are thankful to Mr. Prateek Kothari and Mr. G. S. Negi for PECVD deposition and thermal oxide growth, respectively. Dr. Debashree Banerjee is acknowledged for useful discussions.